\documentclass[twocolumn,showpacs,amsmath,amssymb,aps,prc]{revtex4-1}

\usepackage{graphicx}
\usepackage{dcolumn}
\usepackage{bm}

\begin{document}

\title{UrQMD Study of the Effects of Centrality Definitions on Higher Moments of Net Protons at RHIC}

\author{Gary D. Westfall}
\affiliation{%
National Superconducting Cyclotron Laboratory and Department of Physics and Astronomy, Michigan State University, East Lansing, Michigan 48824
}%

\date{\today}

\begin{abstract}
A study using UrQMD is presented concerning the higher moments of net protons from Au+Au collisions at 7.7, 11.5, 14.6, 19.6, 27, 39, 62.4, and 200 GeV, concentrating on $C_{4}/C_{2} = \kappa\sigma^{2}$.  Higher moments of net protons are predicted to be a sensitive probe of the critical point of QCD.  At the QCD critical point, particular ratios of the moments of net protons are predicted to differ from the Poisson baseline.  Recently STAR has published the higher moments of net protons for Au+Au collisions at $\sqrt{s_{\rm NN}}$ = 7.7, 11.5, 19.6, 27, 39, 62.4, and 200 GeV.  UrQMD quantitatively reproduces STAR's measured $C_{4}/C_{2} = \kappa\sigma^{2}$ for net protons for all Au+Au collisions more central than 30\% and at all centralities for $\sqrt{s_{\rm NN}}$ = 7.7 and 11.5 GeV.  The effects are investigated of three different centrality definitions on the values of $C_{4}/C_{2} = \kappa\sigma^{2}$ from UrQMD calculations including the impact parameter given by UrQMD.  It is shown that using a centrality definition based on multiplicity to calculate the higher moments of net protons gives a biased answer for $C_{4}/C_{2} = \kappa\sigma^{2}$, except in the most central bin (0-5\%).
\end{abstract}

\pacs{25.75.Gz}
\maketitle


\section{Introduction}
Relativistic heavy ion collisions are an essential experimental tool for studying the transition of hadronic matter to partonic matter.  The goal of these experiments is to study the quark gluon plasma (QGP) and the phase diagram of quantum chromodynamics (QCD) \cite{STARWhitePaper}.  The QCD phase diagram is often represented in terms of the baryonic chemical potential ($\mu_{\rm B}$) and the temperature ($T$) \cite{ChemicalFreezeOut}.  The transition from hadronic to partonic matter in the QCD phase diagram is often represented as a line in the QGP phase diagram with hadronic matter existing at low $T$ and low $\mu_{\rm B}$ and quark-gluon matter existing at high $T$ and high $\mu_{\rm B}$.  Lattice QCD calculations (LQCD) indicate a smooth crossover between hadronic matter and quark gluon matter as a function of $T$ at $\mu_{\rm B} \approx 0$ and a first order phase transition as a function of $T$ at large $\mu_{\rm B}$ \cite{SmoothCrossOverAoki,FirstOrderCrossOverEjiri,QCDCheng,QCDFiniteMuB}.  A critical point (CP) in the QCD phase diagram could occur where the first order crossover ends \cite{QCDFiniteMuB,CriticalPointsKapusta,CriticalPointsFodorKatz,CriticalPointsStephanov1,CriticalPointsStephanov2,CriticalPointsStephanov3,CriticalPointsStephanov4,CriticalPointsBleicher}.  Various regions in the QCD phase diagram can be explored using specific Au+Au collision energies.  The Relativistic Heavy Ion Collider (RHIC) at Brookhaven National Laboratory has carried out a Beam Energy Scan (BES) \cite{BESWhitePaper} with the intent of studying regions of the QCD phase diagram at low $\mu_{\rm B}$ ($\sqrt{s_{\rm NN}}$ = 200 GeV) through regions of high $\mu_{\rm B}$ ($\sqrt{s_{\rm NN}}$ = 7.7 GeV).

A possible signal of the CP is non-monotonic behavior of observables related to correlations and fluctuations \cite{CriticalPointsStephanov4}.  One such observable is the event-by-event distribution of conserved quantities within a limited acceptance such the net baryon number.  The net baryon number can be represented by the experimentally accessible net proton number defined as the number of protons minus the number of antiprotons per event ($\Delta N_{\rm p} = N_{\rm proton} - N_{\rm antiproton}$) \cite{KitazawaAsakawa1,KitazawaAsakawa2,BzdakKoch}.  Model predictions show that the higher moments of the net proton multiplicity distribution are sensitive to the QCD CP \cite{CriticalPointsStephanov5,CriticalPointsStephanov6}.  In particular the ratio of the fourth cumulant ($C_{4}$) to the second cumulant ($C_{2}$), which is equal to the product of the kurtosis ($\kappa$) times the variance ($\sigma^{2}$), $C_{4}/C_{2} = \kappa\sigma^{2}$, of the net proton multiplicity distribution is predicted to deviate from the Poisson baseline near the CP.

This paper presents a study of the higher moments of the net proton distribution from Au+Au collisions using the UrQMD model.  UrQMD quantitatively reproduces STAR's measured $C_{4}/C_{2} = \kappa\sigma^{2}$ for net protons for all Au+Au collisions more central than 30\%.  UrQMD quantitatively reproduces STAR's measured $C_{4}/C_{2} = \kappa\sigma^{2}$ for net protons for Au+Au collisions at all centralities for $\sqrt{s_{\rm NN}}$ = 7.7 and 11.5 GeV.  It is shown that using a centrality definition based on the multiplicity of charged particles in the same kinematic regime as the protons and antiprotons used to calculate the moments of net protons gives biased results for $C_{4}/C_{2} = \kappa\sigma^{2}$ except for the most central collisions.

\begin{figure*}
\includegraphics[width=5.25in]{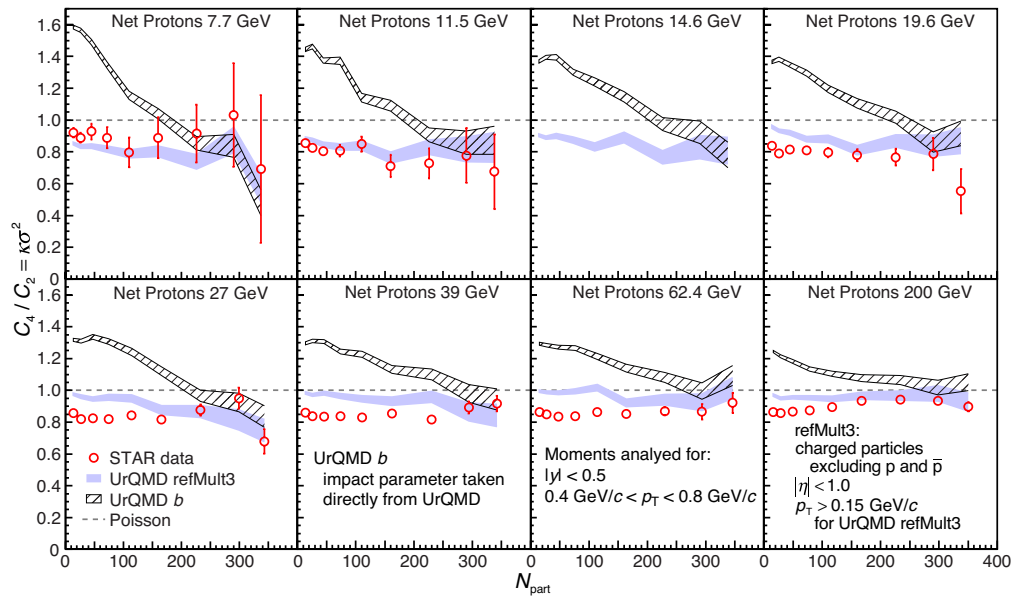}
\caption{\label{fig:fig01}(Color online) Centrality dependence of $C_{4}/C_{2} = \kappa\sigma^{2}$ in Au+Au collisions at $\sqrt{s_{\rm NN}}$ = 7.7, 11.5, 14.6, 19.6, 27, 39, 62.4, and 200 GeV. The symbols represent the STAR data \cite{STARNetProtonPRL}.  The hatched bands represent UrQMD calculations using $b$ from UrQMD to determine the centrality and the solid bands represent UrQMD calculations using refMult3 to determine the centrality.  The dashed lines represent the Poisson baseline.}
\end{figure*}

\section{The UrQMD Model}
Ultra-relativistic Quantum Molecular Dynamics (UrQMD) \cite {UrQMD1,UrQMD2} is a microscopic model used to simulate ultra-relativistic heavy ion collisions at incident energies ranging from 1 AGeV fixed target energies up to collider energies of $\sqrt{s_{\rm NN}}$ = 200 GeV.  UrQMD treats ultra-relativistic heavy collisions in terms of a microscopic transport model that describes the phenomenology of hadronic interactions below $\sqrt{s_{\rm NN}}$ = 5 GeV through the interaction between known hadrons and resonances.  At incident energies above $\sqrt{s_{\rm NN}}$ = 5 GeV, the excitation of color strings and their fragmentation into hadrons dominates the production of particles in UrQMD.  Thus, UrQMD provides an ideal tool with which to study ultra-relativistic heavy ion collisions in the absence of effects from the QCD CP.  UrQMD also enables the study of effects due to correlations and fluctuations of the produced hadrons on specific observables such as the higher moments of net protons from ultra-relativistic heavy ion collisions.  UrQMD version 3.3 with default parameters is used for the results presented in this paper.

\section{Comparison with STAR Data}
STAR recently published \cite{STARNetProtonPRL} the higher moments of the net proton distributions for Au+Au collisions at $\sqrt{s_{\rm NN}}$ = 7.7, 11.5, 19.6, 27, 39, 62.4, and 200 GeV as a function of centrality (the fraction of the total cross section) ranging from 0-5\% (most central) to 70-80\% (most peripheral).  The value of $C_{4}/C_{2} = \kappa\sigma^{2}$ was observed to deviate from the expected Poisson baseline.  However, before these deviations can be ascribed to effects of the CP, effects from known physics must be considered.  For example, in Ref. \cite{STARNetProtonPRL}, STAR presented a sampled-singles model that assumed the independent production of protons and antiprotons that reproduced the measured higher moments of net protons including the deviations from the Poisson baseline.  Here comparisons are made to the STAR data for $C_{4}/C_{2} = \kappa\sigma^{2}$ using UrQMD.

The moments of net protons published by STAR for $C_{4}/C_{2} = \kappa\sigma^{2}$ were calculated using protons and antiprotons with $\left| y \right| < 0.5$ and $0.4 < {p_{\rm{T}}} < 0.8$ GeV/$c$ where $y$ is the rapidity and ${p_{\rm{T}}}$ is the transverse momentum of the protons and antiprotons.  The moments were corrected for efficiency.  The centrality cuts were made using the multiplicity of charged particles measured in the STAR time projection chamber (TPC) \cite{TPCref} with $\left| \eta  \right| < 1$ that were not protons or antiprotons.  This centrality definition is termed refMult3 and was intended to allow the determination of centrality without biasing the measured moments.  To compare with these data, refMult3 was calculated with UrQMD using all charged particles that were not protons or antiprotons that had $\left| \eta  \right| < 1$ and ${p_{\rm{T}}} > 0.15$ GeV/$c$.  

Figure~\ref{fig:fig01} shows the comparison between the STAR data and the UrQMD predictions as a function of centrality for a given incident energy.  Also shown are UrQMD calculations for 14.6 GeV for which there are no data. The calculation of $C_{4}/C_{2} = \kappa\sigma^{2}$ for net protons using UrQMD was done using the same techniques employed by STAR \cite{STARNetProtonPRL}.  For 7.7 GeV, the ${\chi ^2}$ = 13.4 for 9 degrees of freedom and for 11.5 GeV, the ${\chi ^2}$ = 11.9 for 9 degrees of freedom, which means that UrQMD quantitatively reproduces the measured values of  $C_{4}/C_{2} = \kappa\sigma^{2}$ for net protons from these two energies at all centralities.  For the higher energies, the ${\chi ^2}$ is above 250 for 9 degrees of freedom.

\begin{figure*}
\includegraphics[width=5.25in]{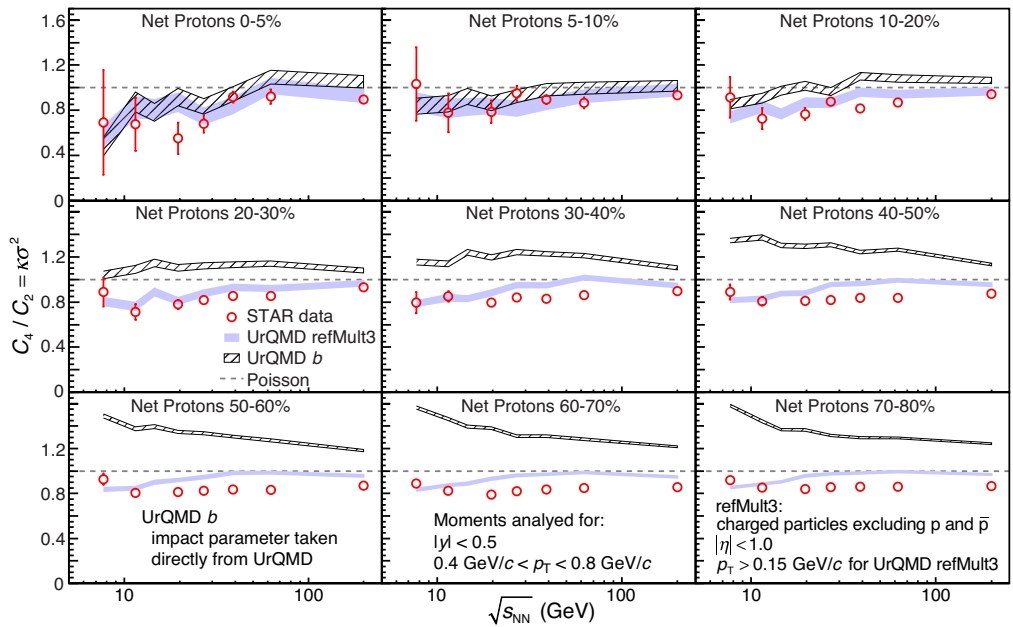}
\caption{\label{fig:fig02}(Color online)  Incident energy dependence of $C_{4}/C_{2} = \kappa\sigma^{2}$ in Au+Au collisions at nine centralities: 0-5\%, 5-10\%, 10-20\%, 20-30\%, 30-40\%, 40-50\%, 50-60\%, 60-70\%, and 70-80\%.  The symbols represent the STAR data \cite{STARNetProtonPRL}. The hatched bands represent UrQMD calculations using $b$ from UrQMD to determine the centrality and the solid bands represent UrQMD calculations using refMult3 to determine the centrality.  The dashed lines represent the Poisson baseline. }
\end{figure*}

Figure~\ref{fig:fig02} shows the comparison between the STAR data and the UrQMD predictions as a function of incident energy for a given centrality.  Also shown are UrQMD calculations for 14.6 GeV for which there are no data. The ${\chi ^2}$ = 13.5, 3.1, 7.9, and 13.1 for 7 degrees of freedom for 0-5\%, 5-10\%, 10-20\%, and 20-30\% respectively.  This comparison means that UrQMD quantitatively reproduces the measured values of  $C_{4}/C_{2} = \kappa\sigma^{2}$ for net protons for all centralities from 0 - 30\% for all incident energies.  For the more peripheral centralities, the ${\chi ^2}$ is above 80 for 7 degrees of freedom.

\begin{figure*}
\includegraphics[width=5.25in]{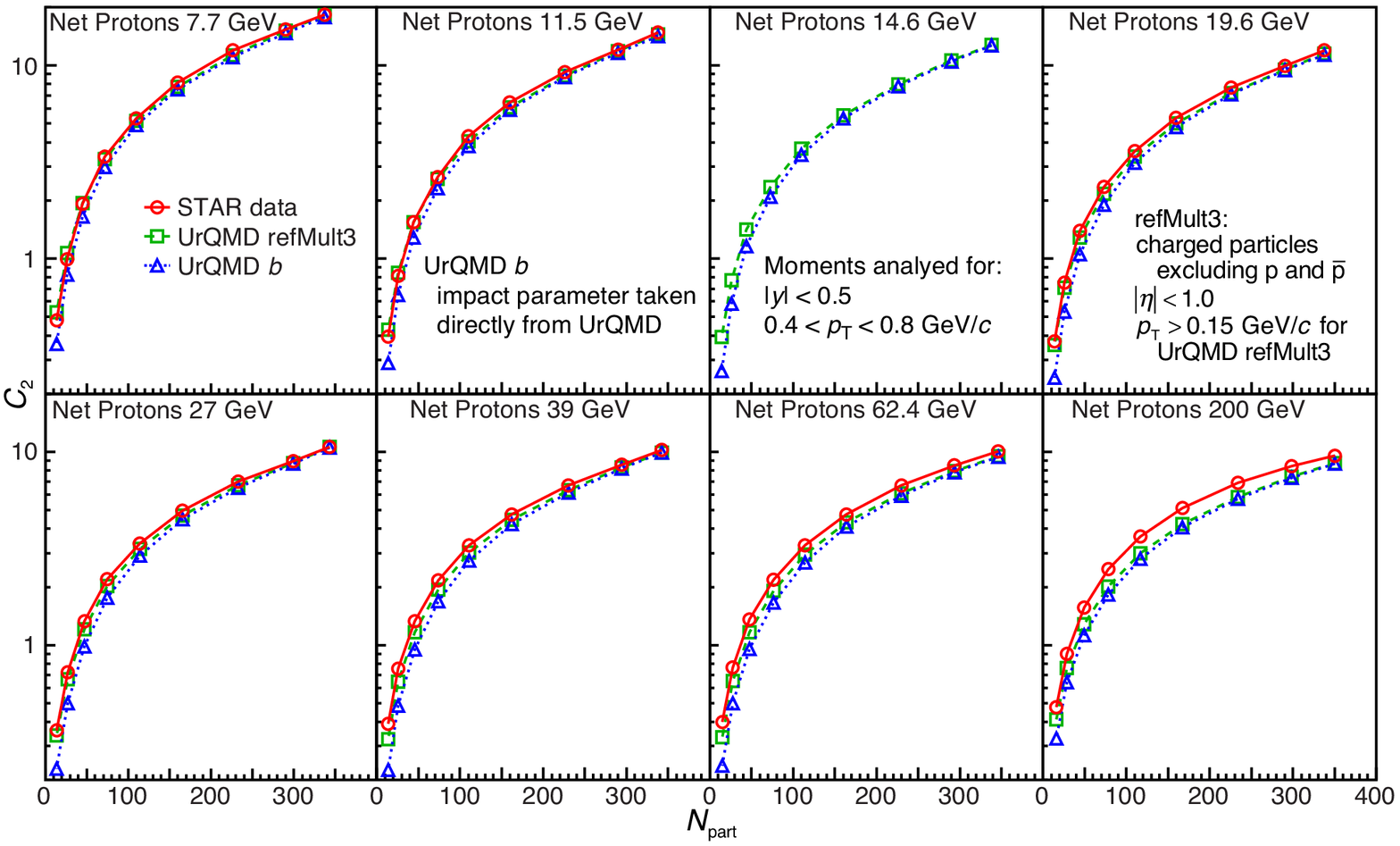}
\caption{\label{fig:fig03}(Color online)  Centrality dependence of UrQMD calculations for $C_{2}$ for net protons using refMult3 and $b$ from UrQMD to determine the centrality for Au+Au collisions at $\sqrt{s_{\rm NN}}$ = 7.7, 11.5, 14.6, 19.6, 27, 39, 62.4, and 200 GeV.  The open circles represent $C_{2}$ for net protons measured by STAR \cite{STARNetProtonPRL}.  Lines are drawn to guide the eye.}
\end{figure*}

\begin{figure*}
\includegraphics[width=5.25in]{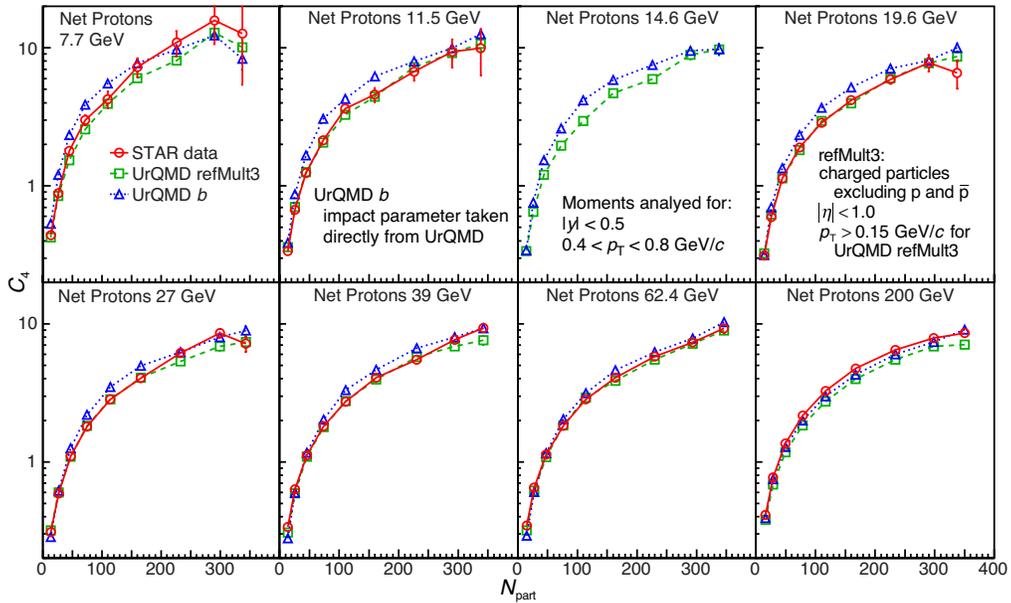}
\caption{\label{fig:fig04}(Color online)  Centrality dependence of UrQMD calculations for $C_{4}$ for net protons using refMult3 and $b$ from UrQMD to determine the centrality for Au+Au collisions at $\sqrt{s_{\rm NN}}$ = 7.7, 11.5, 14.6, 19.6, 27, 39, 62.4, and 200 GeV.  The open circles represent $C_{4}$ for net protons measured by STAR \cite{STARNetProtonPRL}.  Lines are drawn to guide the eye.}
\end{figure*}

\section{Centrality Selection}
One difficulty with using refMult3 as the centrality definition when calculating the higher moments of net protons is that the protons and antiprotons used to calculate the higher moments are correlated with the particles used to define the centrality.  For example, protons and antiprotons are strongly correlated with pions \cite{ParticleRatioFluctuations,HuiWangThesis} in this kinematic region.  These correlations could reduce $C_{2}$ and $C_{4}$.  Another difficulty with using refMult3 to determine the centrality is that there may be volume fluctuations that smear the observed number of protons and antiprotons for a given centrality, which could increase $C_{2}$ and $C_{4}$. To help interpret the effect of using refMult3 on the higher moments of net protons, the impact parameter, $b$, from UrQMD is used to determine the centrality, which should provide a benchmark for what would be expected for the higher moments of net protons.

\begin{figure*}
\includegraphics[width=5.25in]{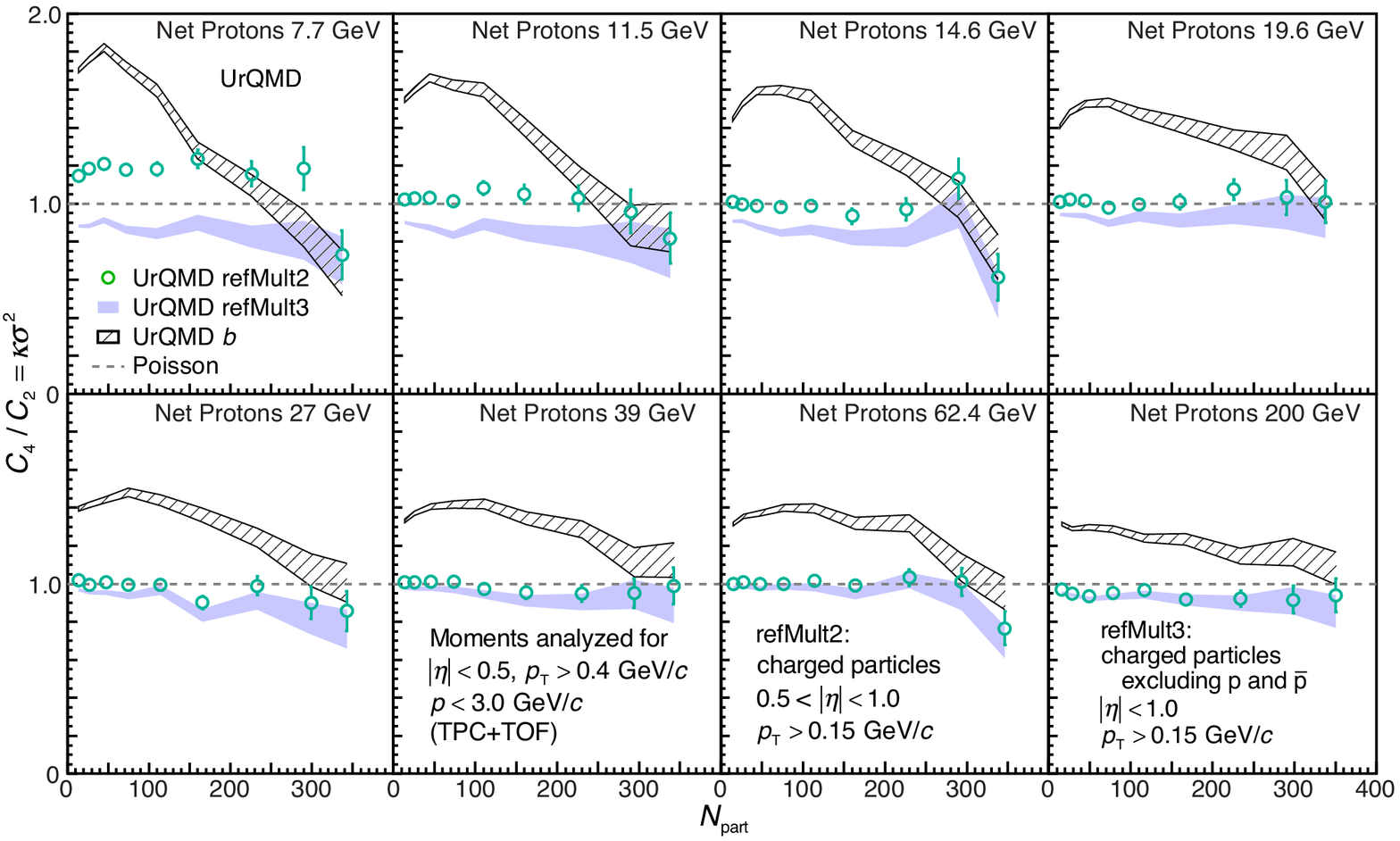}
\caption{\label{fig:fig05}(Color online)  Centrality dependence of $C_{4}/C_{2} = \kappa\sigma^{2}$ in Au+Au collisions at $\sqrt{s_{\rm NN}}$ = 7.7, 11.5, 14.6, 19.6, 27, 39, 62.4, and 200 GeV from UrQMD using refMult2, refMult3, and $b$ from UrQMD to determine the centrality.  The dashed lines represent the Poisson baseline.}
\end{figure*}

\begin{figure*}
\includegraphics[width=5.25in]{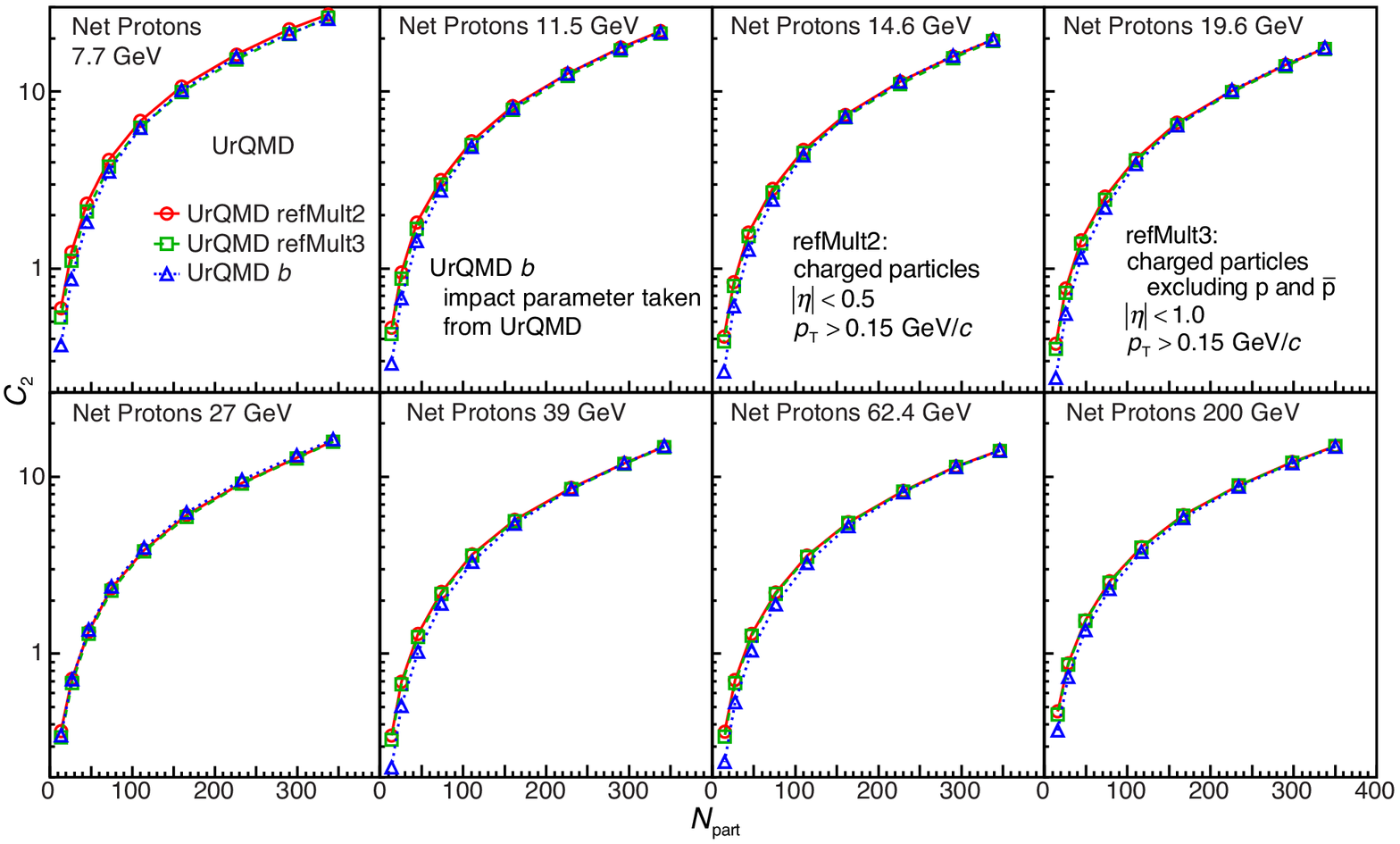}
\caption{\label{fig:fig06}(Color online)  Centrality dependence of $C_{2}$ of net protons in Au+Au collisions at $\sqrt{s_{\rm NN}}$ = 7.7, 11.5, 14.6, 19.6, 27, 39, 62.4, and 200 GeV from UrQMD using refMult2, refMult3, and $b$ from UrQMD to determine the centrality.  Lines are drawn to guide the eye.}
\end{figure*}

Figures \ref{fig:fig01} and \ref{fig:fig02} show the UrQMD predictions using $b$ from UrQMD to determine the centrality.  The values of $C_{4}/C_{2} = \kappa\sigma^{2}$ for net protons are much lower in peripheral collisions when refMult3 is used to determine the centrality than when $b$ from UrQMD is used to determine the centrality.  As discussed above, the UrQMD calculations using refMult3 are close to the data.  In central collisions, the values of $C_{4}/C_{2} = \kappa\sigma^{2}$ for net protons are nearly the same for the two different centrality definitions.   Thus the values of $C_{4}/C_{2} = \kappa\sigma^{2}$ are lowered in peripheral collisions when refMult3 is used as the centrality definition.  The question then arises whether the lowering of $C_{4}/C_{2} = \kappa\sigma^{2}$ when refMult3 is used is due to the lowering of $C_{4}$, the raising of $C_{2}$, or some combination of both.

Figure~\ref{fig:fig03} shows the cumulant $C_{2}$ for net protons as a function of centrality from STAR \cite{STARNetProtonPRL} for Au+Au collisions at $\sqrt{s_{\rm NN}}$ = 7.7, 11.5, 19.6, 27, 39, 62.4, and 200 GeV using refMult3 along with UrQMD calculations for $\sqrt{s_{\rm NN}}$ = 7.7, 11.5, 14.6, 19.6, 27, 39, 62.4, and 200 using refMult3 and $b$ from UrQMD to determine the centrality.  The cumulant $C_{2}$ scales smoothly with the number of participating nucleons, $N_{\rm part}$.  There is little difference between $C_{2}$ for net protons as measured by STAR and the UrQMD calculations of $C_{2}$ using refMult3 except at $\sqrt{s_{\rm NN}}$ = 200 GeV where the values of $C_{2}$ measured by STAR are higher than the values calculated using UrQMD.  In the most peripheral collisions at all incident energies, the values of $C_{2}$ calculated using refMult3 to determine the centrality are higher than the values of $C_{2}$ calculated using $b$ from UrQMD to determine the centrality, which indicates that volume fluctuations could be important in the determination of $C_{2}$.  At centralities more central than 50\%, the values of $C_{2}$ calculated using UrQMD using refMult3 and $b$ from UrQMD are similar.

\begin{figure*}
\includegraphics[width=5.25in]{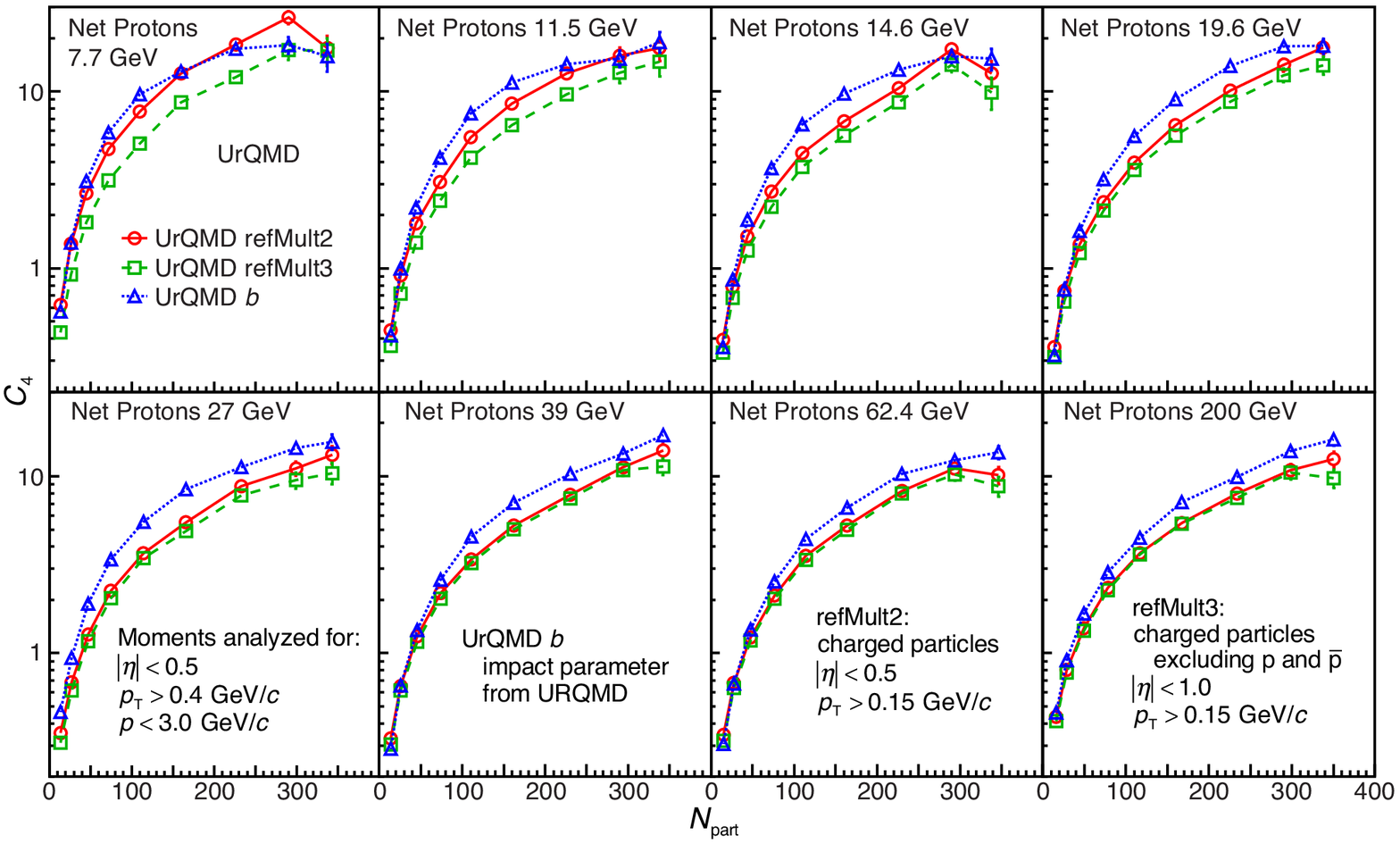}
\caption{\label{fig:fig07}(Color online)  Centrality dependence of $C_{4}$ of net protons in Au+Au collisions at $\sqrt{s_{\rm NN}}$ = 7.7, 11.5, 14.6, 19.6, 27, 39, 62.4, and 200 GeV from UrQMD using refMult2, refMult3, and $b$ from UrQMD to determine the centrality. Lines are drawn to guide the eye.}
\end{figure*}

Figure~\ref{fig:fig04} shows the cumulant $C_{4}$ as a function of centrality from STAR \cite{STARNetProtonPRL} for Au+Au collisions at $\sqrt{s_{\rm NN}}$ = 7.7, 11.5, 19.6, 27, 39, 62.4, and 200 GeV using refMult3 along with UrQMD calculations for $\sqrt{s_{\rm NN}}$ = 7.7, 11.5, 14.6, 19.6, 27, 39, 62.4, and 200 using refMult3 and $b$ from UrQMD to determine the centrality.  The values of $C_{4}$ calculated using UrQMD with refMult3 as the centrality definition agree reasonably well with the values of $C_{4}$ measured by STAR.  However, the values of $C_{4}$ calculated using UrQMD using $b$ from UrQMD as the centrality definition are generally higher except in the most central bin (0-5\%), where the measured values of $C_{4}$ and the values of $C_{4}$ calculated using both centrality definitions agree fairly well.  Also visible in Figure~\ref{fig:fig04} are relatively large fluctuations in $C_{4}$ calculated using UrQMD with refMult3 as the centrality definition in central collisions at incident energies at and below 27 GeV, which are similar to those observed by STAR.

Thus the increases in $C_{4}/C_{2} = \kappa\sigma^{2}$ shown in Figures~\ref{fig:fig01} and~\ref{fig:fig02} in the UrQMD calculation using $b$ from UrQMD as the centrality definition are due to increases in $C_{4}$ while $C_{2}$ remains almost unchanged for collisions more central that 50\%.  For collisions more peripheral than 50\%, the increase in UrQMD calculations for $C_{4}/C_{2} = \kappa\sigma^{2}$ using $b$ from UrQMD as the centrality definition is due an increase in $C_{4}$ along with a decrease in $C_{2}$.  

One reason for the decrease in the values of $C_{4}$ from UrQMD when refMult3 is used compared to the values of $C_{4}$ when $b$ from UrQMD is used could be that there are correlations between the protons and antiprotons used to calculate moments of net protons and the particles included in the refMult3 centrality definition.   A possible approach to removing the effects of the centrality definition refMult3 would be to use a centrality definition that does not include particles that overlap kinematically with the protons and antiprotons used to calculated the moments of net protons.  STAR used this method for the higher moments of net charge \cite{STARNetChargePRL}.  The procedure involved using charged particles with $\left| \eta  \right| < 0.5$ to calculate the higher moments of net charge while using charged particles with $0.5 < \left| \eta  \right| < 1.0$ to determine the centrality.  Thus the particles used to calculate the moments did not overlap kinematically with the particles used to define the centrality.  This centrality definition is termed refMult2.

Figure~\ref{fig:fig05} shows a similar analysis for the higher moments of net protons using UrQMD.  The higher moments of net protons are calculated with $\left| \eta  \right| < 0.5$ with $p_{\rm T} >$ 0.4 GeV/$c$ and $p <$ 3.0 GeV/$c$ corresponding to the the momentum acceptance of the STAR TPC using time-of-flight information (TPC+TOF) \cite{TPCref,TOFref}.  The centrality is determined from UrQMD using refMult2, refMult3, and $b$ from UrQMD.  The centrality selection criterion has little effect on $C_{4}/C_{2} = \kappa\sigma^{2}$ in the most central bin.  At centralities other than 0-5\%, the values of $C_{4}/C_{2} = \kappa\sigma^{2}$ where the centrality was selected using $b$ from UrQMD are higher than for the two multiplicity-based centrality definitions.  At the higher incident energies, the values of $C_{4}/C_{2} = \kappa\sigma^{2}$ calculated using refMult2 and refMult3 are similar.  At the lower incident energies, the values of $C_{4}/C_{2} = \kappa\sigma^{2}$ calculated using refMult2 are considerably higher than the values of $C_{4}/C_{2} = \kappa\sigma^{2}$ calculated using refMult3.  The question again arises whether the lowering of $C_{4}/C_{2} = \kappa\sigma^{2}$ when refMult2 and refMult3 are used is due to the lowering of $C_{4}$, the raising of $C_{2}$, or some combination of both.

Figure~\ref{fig:fig06} shows the cumulant $C_{2}$ for net protons as a function of centrality calculated using UrQMD for Au+Au collisions at$\sqrt{s_{\rm NN}}$ = 7.7, 11.5, 14.6, 19.6, 27, 39, 62.4, and 200 using refMult2, refMult3 and $b$ from UrQMD to determine the centrality.  There is little difference between $C_{2}$ for net protons calculated using UrQMD using refMult2 and refMult3 as the centrality definition.  In the most peripheral collisions at all incident energies, the values of $C_{2}$ calculated using refMult2 and refMult3 to determine the centrality are higher than the values of $C_{2}$ calculated using $b$ from UrQMD to determine the centrality, which indicates that volume fluctuations could be important in the determination of $C_{2}$.  At centralities more central than 50\%, the values of $C_{2}$ calculated using UrQMD using refMult2, refMult3, and $b$ from UrQMD are similar.

Figure~\ref{fig:fig07} shows the cumulant $C_{4}$ as a function of centrality calculated using UrQMD for Au+Au collisions at $\sqrt{s_{\rm NN}}$ = 7.7, 11.5, 14.6, 19.6, 27, 39, 62.4, and 200 using refMult2, refMult3 and $b$ from UrQMD to determine the centrality.  At the lower incident energies, the values of $C_{4}$ calculated using refMult2 as the centrality definition are higher than the values of $C_{4}$ calculated using refMult3 as the centrality definition.  At the higher energies, the values of $C_{4}$ calculated using refMult2 as the centrality definition are comparable to the values of $C_{4}$ calculated using refMult3 as the centrality definition.  At all incident energies, the values of $C_{4}$ calculated using $b$ from UrQMD as the centrality definition are higher that the values calculated using refMult2 or refMult3 as the centrality definition.  Also visible in Figure~\ref{fig:fig07} are relatively large fluctuations in $C_{4}$ calculated using UrQMD with refMult2 and refMult3 as the centrality definition in central collisions, which are similar to those observed by STAR.

At $\sqrt{s_{\rm NN}}$  = 7.7 and 11.5 GeV, the values of $C_{4}$ calculated using refMult2 as the centrality definition are markedly higher that the values of $C_{4}$ calculated using refMult3 for non-central collisions.  This is also reflected in $C_{4}/C_{2} = \kappa\sigma^{2}$ at 7.7 and 11.5 GeV as shown in Figure~\ref{fig:fig04}.  The correlations between protons and antiprotons used to calculate the moments and the particles used to determine the centrality with refMult3 get stronger at lower incident energies and in peripheral collisions \cite{ParticleRatioFluctuations,HuiWangThesis}.  These correlations would make $C_{4}$ smaller when refMult3 is used to determine the centrality.  On the other hand, the number of particles included in refMult2 is about half the number of particles included in refMult3 and volume fluctuations may increase $C_{4}$ when refMult2 is used to determine the centrality in non-central collisions.

\section{Conclusions}

UrQMD quantitatively reproduces STAR's measured $C_{4}/C_{2} = \kappa\sigma^{2}$ for net protons for Au+Au collisions more central than 30\% and STAR's measured $C_{4}/C_{2} = \kappa\sigma^{2}$ for net protons for Au+Au collisions at all centralities for $\sqrt{s_{\rm NN}}$ = 7.7 and 11.5 GeV.  The centrality definition affects the determination of $C_{4}/C_{2} = \kappa\sigma^{2}$ for net protons in Au+Au collisions at RHIC at all centralities except for the most central collisions.  Using refMult3, which is based on multiplicity, to determine the centrality lowers the measured values of $C_{4}/C_{2} = \kappa\sigma^{2}$ by lowering the cumulant $C_{4}$ for collisions more central that 50\%.  For collisions more peripheral than 50\%, the lowering of the measured values of $C_{4}/C_{2} = \kappa\sigma^{2}$ when using refMult3 as the centrality definition occurs because $C_{4}$ is lowered and $C_{2}$ is raised.  When $b$ from UrQMD is used to determine the centrality, the values of $C_{4}/C_{2} = \kappa\sigma^{2}$ are larger and better represent the actual higher moments of the net protons.  Care should be taken in comparing the results for higher moments of net protons to theoretical models at any centrality other than the most central centrality when multiplicity-based centrality definitions such as refMult3 are used.

\begin{acknowledgements}
The author thanks E. Sangaline and W.J. Llope for fruitful discussions and inspiration.  This work is supported by DOE Grant No. DE-FG02-98ER41070.
\end{acknowledgements}.


\begin{thebibliography}{99}

\bibitem{STARWhitePaper}
J. Adams {\em et al.} (STAR Collaboration), Nucl. Phys. A {\bf 757}, 102 (2005).

\bibitem{ChemicalFreezeOut}
J. Cleymans, H. Oeschler, K. Redlich, and S. Wheaton,
J. Phys. G: Nucl. Part. Phys. {\bf 32}, S165 (2006).

\bibitem{SmoothCrossOverAoki}
Y. Aoki, G. Endrodi, Z. Fodor, S.D. Katz, and K.K. Szab\'o, 
Nature {\bf 443}, 675 (2006).

\bibitem{FirstOrderCrossOverEjiri}
S. Ejiri, Phys. Rev. D {\bf 78}, 074507 (2008).

\bibitem{QCDCheng}
M. Cheng {\em et al.},
Phys. Rev. D {\bf 77}, 014511 (2008).

\bibitem{QCDFiniteMuB}
R.V. Gavai and S. Gupta,
Phys. Rev. D {\bf 78}, 114503 (2008).

\bibitem{CriticalPointsKapusta}
E.S. Bowman and J.I. Kapusta,
Phys. Rev. C {\bf 79}, 015202 (2009).

\bibitem{CriticalPointsFodorKatz}
Z. Fodor and S.D. Katz,
JHEP 0404, 050 (2004).

\bibitem{CriticalPointsStephanov1}
M. Stephanov,
Int. J. Mod. Phys. A {\bf 20}, 4387 (2005).

\bibitem{CriticalPointsStephanov2}
M. Stephanov,
Prog. Theor. Phys. Supplement {\bf 153}, 139 (2004).

\bibitem{CriticalPointsStephanov3}
M. Stephanov, K. Rajagopal, and E. Shuryak,
Phys. Rev. Lett. {\bf 81}, 4816 (1998).

\bibitem{CriticalPointsStephanov4}
M. Stephanov, K. Rajagopal, and E. Shuryak,
Phys. Rev. D {\bf 60}, 114028 (1999).

\bibitem{CriticalPointsBleicher}
C. Herold, M. Nahrgang, I. Mishustin, and M. Bleicher,
Nucl. Phys. A {\bf 925}, 14 (2014).

\bibitem{BESWhitePaper}
M.M. Aggarwal {\em et al.} (STAR Collaboration), arXiv:1007.2613.

\bibitem{KitazawaAsakawa1}
M. Kitazawa and M. Asakawa,
Phys. Rev. C {\bf 86}, 024904 (2012).

\bibitem{KitazawaAsakawa2}
M. Kitazawa and M. Asakawa,
Phys. Rev. C {\bf 86}, 069902(E) (2012).

\bibitem{BzdakKoch}
A. Bzdak and V. Koch,
Phys. Rev. C {\bf 86}, 044904 (2012).

\bibitem{CriticalPointsStephanov5}
C. Athanasiou, K. Rajagopal, and M. Stephanov,
Phys. Rev. D {\bf 82}, 074008 (2010).

\bibitem{CriticalPointsStephanov6}
M. A. Stephanov,
Phys. Rev. Lett. {\bf 102}, 032301 (2009).

\bibitem{UrQMD1}
S.A. Bass {\em et al.},
Prog. Part. Nucl. Phys. {\bf 41}, 225 (1998).

\bibitem{UrQMD2}
H. Petersen, J. Steinheimer, G. Berau, M. Bleicher, and H. St\"{o}cker,
Phys. Rev. C {\bf 78}, 044901 (2008).

\bibitem{STARNetProtonPRL}
L. Adamczyk {\em et al.} (STAR Collaboration),
Phys. Rev. Lett. {\bf 112}, 032302 (2014).

\bibitem{TPCref}
K.H. Ackermann {\it et al.}, (STAR Collaboration),
Nucl. Instrum. Methods A {\bf 499}, 624 (2003).

\bibitem{ParticleRatioFluctuations}
N.M. Abdelwahab {\em et al.} (STAR Collaboration), arXiv:11410.5375.

\bibitem{HuiWangThesis}
H. Wang, Ph.D. Thesis, Michigan State University (2012).

\bibitem{STARNetChargePRL}
L. Adamczyk {\em et al.} (STAR Collaboration),
Phys. Rev. Lett. {\bf 113}, 092301 (2014).

\bibitem{TOFref}
B. Bonner {\it et al.}, Nucl. Instrum. Methods A {\bf 508}, 181 (2003).

\end{thebibliography}
\end{document}